\documentclass[twocolumn,showpacs,preprintnumbers,prb,amsmath,amssymb,floatfix]{revtex4}

\usepackage{epsfig,psfrag}
\usepackage{dcolumn}
\usepackage{bm}
\usepackage{graphicx}
\usepackage{color}

\begin{document}

\title{Vibration-induced correction to the current through a single molecule} 
\author{R.~Egger$^1$ and A.O.~Gogolin$^{1,2}$}
\affiliation{$^1$~Institut f\"ur Theoretische Physik,
Heinrich-Heine-Universit\"at D\"usseldorf, D-40225  D\"usseldorf, Germany\\
$^2$~Department of Mathematics, Imperial College London, 180 Queen's Gate,
London SW7 2AZ, United Kingdom}

\date{\today}

\begin{abstract}
We provide analytical results for the perturbative correction
to the current-voltage relation through a vibrating molecule 
for weak electron-phonon coupling.  The nonlinear conductance 
exhibits a steplike feature at $eV=\hbar \omega_0$, where
$\omega_0$ is the vibration frequency. 
We establish criteria for the sign change 
of the step in the conductance (up or down).
This transition turns out to be nonuniversal and 
is  governed by essentially all system parameters.
\end{abstract}
\pacs{73.23.-b, 73.63.-b, 72.10.Di }

\maketitle
Electronic transport through single molecules has attracted much 
attention lately; for recent reviews,
see Refs.~\onlinecite{nitzan,book,tsukada,tao,nitzan2}.
Besides the technological promises raised by molecular electronics,
this field also poses interesting questions to theory.
In this Brief Report, we will revisit the problem of 
how the coupling to a vibrational mode (``phonon'') of frequency
$\omega_0$ and electron-phonon coupling strength $g$ 
affects the current through a single molecule.   
The resulting features in the $I$-$V$ characteristics are
often referred to as inelastic electron tunneling spectroscopy.\cite{nitzan2} 
Theories for various aspects of this rich and diverse
problem have been proposed over the past few 
years by a large number of 
authors,\cite{schoeller,flensberg,mitra,paulsson,cornaglia,alfredo,milena,zazu}
primarily motivated by groundbreaking experiments demonstrating the influence
of vibrational degrees of freedom in single-molecule transport.
To mention just a few key experiments, 
single-molecule transport has been
studied using various organic molecules,\cite{organic}   
fullerenes,\cite{park,natelson,paul} carbon nanotubes,\cite{leroy,sapmaz} 
or single hydrogen molecules between Pt leads.\cite{ruitenbeek}
Phonon-assisted processes were shown to
imply a step in the $I$-$V$ characteristics once the dc bias $V$ reaches
the threshold value $\hbar \omega_0/e$ for excitation of 
a phonon mode.  Related features can sometimes be seen at integer
multiples of this value.
Vibrational effects on single-molecule transport have
recently been reviewed,\cite{nitzan2} including a discussion
of the validity regime for our Hamiltonian below. 

It is remarkable that the
experimentally observed vibration-induced step features in the
differential conductance can either
{\sl decrease}\cite{ruitenbeek} or {\sl increase}
the conductance through the molecule
near $eV=\hbar \omega_0$.\cite{organic,park,paul,leroy}
This corresponds to dips versus peaks in the second derivative, $d^2 I/dV^2$.
Such features have also been reported
theoretically.\cite{nitzan2,ueba,nitzan3}
Recent theoretical work on this question\cite{paulsson,alfredo} argues that the
transparency ${\cal T}$ of the single-molecule junction 
basically determines the step direction.
The critical value
was reported to be ${\cal T}=1/2$, with a step down (up) in the differential
conductance at $eV=\hbar \omega_0$
for ${\cal T}>1/2$ (${\cal T}<1/2$).  This conclusion seems  roughly consistent
with existing experimental data: For the $H_2$ measurements,\cite{ruitenbeek} 
${\cal T}$ close to unity was reported, while
typically ${\cal T}\alt 0.1$ in the other experiments.
However,  given the many parameters present in even the simplest Hamiltonian,
one may question why the crossover should be universal in the sense
that it is only determined by the transparency ${\cal T}=1/2$.
Here, we reexamine the question of current increase or decrease
at the phonon excitation threshold $eV=\hbar \omega_0$.
 We derive and discuss analytical results 
for the current correction $\delta I$ perturbative in the electron-phonon
coupling strength $g$. Experimental values for $g$ are often very small,
justifying a truncation of the perturbation series already at the 
lowest nontrivial order.
For that reason, our expressions below are expected to provide useful
estimates for many experiments.  
However, we will not attempt a detailed description of 
specific experiments,  but instead aim at an {\sl analytical}\
understanding of vibrational features in the $I$-$V$ 
characteristics under a simple yet realistic model. 
In our opinion, a thorough 
understanding of the lowest-order
feature is worthwhile, given the complexity of the physical
processes involved.  Most available results are
obtained from lengthy numerical calculations and do not easily yield general
 insights.  In addition, many calculations were based on 
essentially uncontrolled approximations, rendering their
predictive power questionable. Moreover, some published theoretical work on this
subject have used approximate schemes that are in conflict
with the basic requirement of current conservation.
Current conservation is automatically fulfilled under 
self-consistent approximations, but is  generally violated otherwise
(unless particle-hole symmetry is present), limiting the
practical usefulness of many approximations to special 
parameter sets.\cite{hershfield}
The lowest-order correction, however, can be evaluated exactly 
and, therefore, does not suffer from any such limitations.
Although aspects of the lowest-order correction $\delta I$ have been
studied before,\cite{nitzan2,ueba,nitzan3} 
a complete and analytical discussion was not given so far.
We show that besides the step feature caused by inelastic processes,
quasielastic electron-phonon interactions
are responsible for another singular
term that logarithmically diverges at $eV=\hbar \omega_0$.

We study the model of just one relevant molecular level
(``dot''), the so-called local Holstein model, 
also employed by previous work,\cite{nitzan2}
\begin{eqnarray}\label{model}
H&=& (\epsilon_0 +gQ)d^\dagger d  +  \hbar \omega_0 b^\dagger b+
\\ \nonumber &+&  \sum_{k,\alpha=L/R=\pm} ( \gamma_{\alpha}
d^\dagger  c^{}_{k\alpha} + {\rm H.c.}) + \sum_{k,\alpha} 
[\epsilon_{k}-\mu_\alpha] c_{k\alpha}^\dagger c^{}_{k\alpha},
\end{eqnarray}
which neglects the Coulomb interaction $U$ and is formulated
for spinless dot fermion operator $d$ (we set $\hbar=1$).  
The effect of the lowest-order correction in $U$, consistent
with our $g^2$ calculation below, is anyway trivial and can be
absorbed by a renormalization of the chemical potential.
We take the standard wide-band limit
for the leads, which is justified if the lead density
of states $\nu_0$ does not vary significantly in energy on the relevant 
scales.  The lead modes are occupied according to Fermi functions 
$f_{L/R}(\epsilon)=f(\epsilon-\mu_{L/R})$, where $\mu_L-\mu_R=eV$ defines
the bias voltage $V$.
We introduce the hybridizations as
 $\Gamma_{L,R}= \pi \nu_0 |\gamma_{L/R}|^2$, and define $\Gamma=\Gamma_L
+ \Gamma_R$.
Finally, the boson operator $b$ describes an Einstein phonon mode
(vibration mode) of frequency $\omega_0$, with linear coupling
of strength $g$ between the displacement operator
$Q=b+b^\dagger$ and  the dot occupation operator $d^\dagger d$.
The electrical current through the dot can be computed 
from the retarded dot Green's function (GF) evaluated
in the presence of the leads and the phonon, $G^r(\omega)$, according to the
well-known expression \cite{mw}
\begin{equation}\label{cur}
I(V) =- \frac{4e}{h} \frac{\Gamma_L  \Gamma_R}{\Gamma} 
\int d\omega [f_L(\omega)-f_R(\omega)] {\rm Im} G^r(\omega).
\end{equation}
The non-interacting ($g=0$) Keldysh GF describing the 
out-of-equilibrium dot coupled to the leads is
\begin{eqnarray}\label{g0}
\hat G_0(\omega) & =& \frac{1}{(\omega-\epsilon_0)^2+\Gamma^2}  
\Biggl [ (\omega-\epsilon_0) {\rm diag}(1,-1)
\\ \nonumber && -i \sum_\alpha \Gamma_\alpha 
\left( \begin{array}{cc}
 2 f_\alpha(\omega)-1 & -2f_\alpha(\omega) \\
2-2f_\alpha(\omega) & 2f_\alpha(\omega)-1 \end{array} \right)\Biggr].
\end{eqnarray}
Note that we use the unrotated Keldysh notation,
where the connection to retarded or advanced GFs $G^{r/a}$ and
the lesser GF $G^<$ is established by 
\[
\left( \begin{array}{cc} G^{--} & G^{-+} \\ 
G^{+-} & G^{++} \end{array} \right) = 
\left( \begin{array}{cc}
G^{r} + G^{<} & G^{<} \\
G^{r}- G^a + G^{<} & -G^{a}+G^< \end{array} \right),
\]
such that $G_0^r(\omega)= (\omega-\epsilon_0+i\Gamma)^{-1}$.
Since $G^r$ obeys its own Dyson equation, we only need
to compute the retarded self-energy $\Sigma^r(\omega)$
to order $g^2$, resulting in
\begin{equation}\label{dyson}
G^r(\omega) = G_0^r(\omega)+ G_0^r(\omega) \Sigma^r(\omega) G_0^r(\omega),
\end{equation}
where the second term defines the correction $\delta I$ in Eq.~(\ref{cur}).
We will focus on the most interesting $T=0$ limit
from now on, where the Fermi function is $f(\omega)=\Theta(-\omega)$.
Defining 
\begin{equation}\label{mubar}
\bar \mu=\frac{\mu_L+\mu_R}{2}-\epsilon_0, \quad 
\bar \mu_{\alpha=L/R=\pm}= \bar \mu \pm eV/2,
\end{equation}
the first ($g=0$) term yields 
\begin{equation}\label{i0}
I_0(V) =  \frac{e}{h}\frac{4\Gamma_L\Gamma_R}{\Gamma}\left[
\tan^{-1}(\bar \mu_L/\Gamma)-\tan^{-1}(\bar \mu_R/\Gamma)\right].
\end{equation}
The $V\to 0$ transparency  of the
junction, ${\cal T}=(h/e^2) dI/dV$, follows as
\begin{equation} \label{trans}
{\cal T} = \frac{4\Gamma_L \Gamma_R}{\Gamma^2} 
\frac{1}{1+(\bar \mu/\Gamma)^2} \leq 1. 
\end{equation}
Note that $\epsilon_0$ and the mean chemical potential always
appear in the combined scale $\bar\mu$.  

Let us now analyze the lowest-order correction $\delta I$ to the
current.  It arises from the retarded self-energy $\Sigma^r(\omega)$
evaluated to order $g^2$ due to phonon processes, entering
Eqs.~(\ref{dyson}) and (\ref{cur}).  There are two contributions
coming from the real (imaginary) parts $\Sigma_R^r$ ($\Sigma_I^r$),
corresponding to quasielastic (inelastic) processes,
\begin{eqnarray}\label{iqe}
\delta I_{\rm qel} &=& \frac{e}{h}\frac{4\Gamma_L\Gamma_R}{\Gamma}
\int_{\bar\mu_R}^{\bar\mu_L} d\omega \frac{2\omega \Gamma}{(\omega^2
+\Gamma^2)^2} \Sigma_R^r(\omega),\\ \label{ine}
\delta I_{\rm inel} &=& \frac{e}{h}\frac{4\Gamma_L\Gamma_R}{\Gamma}
\int_{\bar\mu_R}^{\bar\mu_L} d\omega \frac{\Gamma^2-\omega^2}{(\omega^2
+\Gamma^2)^2} \Sigma_I^r(\omega).
\end{eqnarray}
The self-energy is readily computed on the perturbative level,
where two diagrams are present in order $g^2$.
The ``tadpole'' diagram does not carry frequency dependence and
can be absorbed by a renormalization of $\bar \mu$. 
We therefore keep only the standard ``rainbow'' diagram,
which gives the retarded self-energy 
$\Sigma^r(\omega)=\Sigma^{--}(\omega)+\Sigma^{-+}(\omega)$
from
\begin{equation}
\Sigma^{-\pm}(\omega) = \mp ig^2 \int \frac{d\Omega}{2\pi} D_0^{-\pm}(\Omega)
G_0^{-\pm} (\omega-\Omega).
\end{equation}
Here, $\hat D_0(\omega)$ is the $g=0$ GF of the displacement operator $Q$,
which for $T=0$ is given by 
\begin{equation}
\hat D_0 (\omega) = 
\left( \begin{array}{cc}
\sum_{s=\pm} \frac{1}{s\omega-\omega_0+i0^+} &
 -2\pi i \delta(\omega+\omega_0) \\
 -2\pi i\delta( \omega-\omega_0)&  -\sum_s \frac{s}{s\omega-\omega_0+i0^+} 
          \end{array} \right).
\end{equation}
Using a Wiener-Hopf decomposition of $G_0^{--}(\omega)$ into the parts
analytic in the upper or lower complex frequency plane, one 
arrives at the result (cf.~also Ref.~\onlinecite{flensberg}),
\begin{eqnarray}\nonumber
\Sigma_R^r(\omega) &=& \sum_{\alpha,s=\pm} \frac{g^2 \Gamma_\alpha}{
\Gamma^2+(\omega+s\omega_0)^2} \Biggl [ \frac{s}{\pi}
\ln \frac{\sqrt{\Gamma^2+\bar \mu_\alpha^2}}{|\omega+s\omega_0-\bar\mu_\alpha|}
\\  \label{sr}
&+& \frac{\omega+s\omega_0}{2\Gamma} \left( 1+\frac{2s}{\pi}\tan^{-1}(\bar 
\mu_\alpha/\Gamma) \right)\Biggr], \\  \label{si}
\Sigma_I^r(\omega) &=& - \sum_{\alpha,s} \frac{g^2 \Gamma_\alpha
\Theta[s(\bar\mu_\alpha-\omega)-\omega_0]}{(\omega+s\omega_0)^2 + \Gamma^2}.
\end{eqnarray}
The computation of $\delta I$ is then reduced to a single frequency integration.
We see that due to the phonon mode, the retarded electron self-energy
contains directly the Fermi functions shifted by $\pm\omega_0$.
Hence, a singular (step) dependence of its imaginary part on
the energy results, which must be accompanied by a logarithmic
singularity in the
real part due to analytic properties. These singularities
have  been discovered first by Engelsberg and Schrieffer 
in their study of bulk Einstein phonons.\cite{engelsberg}
With the above self-energies, it is easy to check that the requirement
for current conservation,\cite{hershfield}
\[
\int d \omega\left[G^<(\omega)\Sigma^>(\omega)-G^>(\omega)
\Sigma^<(\omega)\right]=0,
\]
is fulfilled (also at finite $T$) to the required $g^2$ order.  
Let us first discuss the inelastic part, $\delta I_{\rm inel}$. 
Using the auxiliary relation for an arbitrary function $F(\omega,V)$,
\begin{equation} \label{aux}
\frac{d}{dV}\int_{\bar\mu_R}^{\bar\mu_L} d\omega F(\omega,V)
= \frac{e}{2} \sum_\alpha F(\bar\mu_\alpha,V)+\int_{\bar\mu_R}^{\bar \mu_L}
d\omega \frac{\partial F(\omega,V)}{\partial V},
\end{equation}
some algebra gives the $g^2$ inelastic 
correction to the $T=0$ nonlinear conductance  for arbitrary parameters
 in closed form,
\begin{eqnarray}
\label{nonlcond1}
\frac{d\delta I_{\rm inel}}{dV} &=& - \frac{e^2}{h} \Theta(V-\hbar\omega_0/e)
g^2\frac{2\Gamma_L\Gamma_R}{\Gamma}
\sum_{\alpha} \Gamma_\alpha  \\ \nonumber &\times & \Biggl( 
\frac{\Gamma^2-\bar\mu_{-\alpha}^2}{(\Gamma^2+\bar\mu_{-\alpha}^2)^2
[(\bar \mu_{-\alpha}+\alpha\omega_0)^2 +\Gamma^2]} 
\\ \nonumber &+& 
\frac{\Gamma^2-(\bar\mu_{\alpha}-\alpha\omega_0)^2}{[\Gamma^2+(\bar\mu_{\alpha}
-\alpha\omega_0)^2]^2 (\bar \mu_\alpha^2 +\Gamma^2)} \Biggr).
\end{eqnarray}
We focus now on the singular contribution at $eV=\hbar\omega_0$, which 
is best illustrated by analyzing $d^2 \delta I/dV^2$. 
Singular terms come from the derivative acting on the Heaviside
function in Eq.~(\ref{nonlcond1}), and produce a delta peak,
\begin{equation}\label{step} 
\left . \frac{d^2 \delta I_{\rm inel}}{dV^2} \right  |_{\rm sing}
= -\frac{e^2}{h} (g/\Gamma)^2 \frac{4\Gamma_R\Gamma_L}{\Gamma^2} 
c_{\rm inel} \delta (V-\hbar\omega_0/e) 
\end{equation}
with the dimensionless coefficient 
\begin{equation}\label{cdef}
c_{\rm inel} =  \frac{1-[(\bar \mu/\Gamma)^2- (\omega_0/2\Gamma)^2]^2 +
2\omega_0(\Gamma_L-\Gamma_R)\bar \mu/\Gamma^3}
{ [1+(\bar \mu/\Gamma+\omega_0/2\Gamma)^2]^2 
 [1+(\bar \mu/\Gamma-\omega_0/2\Gamma)^2]^2},
\end{equation}
which is the main result of this Brief Report.
We will discuss this result below, but first turn
to the quasielastic contribution $\delta I_{\rm qel}$
due to the real part (\ref{sr}) of the self-energy.
Using Eq.~(\ref{aux}), we find again a singular
contribution in the differential conductance at $V=\hbar\omega_0/e$. 
 We obtain the analytical result, valid for 
$V\simeq \hbar \omega_0/e$, 
\begin{equation} \label{qelcorr}
\left . \frac{d\delta I_{\rm qel}}{dV} \right |_{\rm sing}
=- \frac{2}{\pi}
\frac{e^2}{h} (g/\Gamma)^2 \frac{4\Gamma_R\Gamma_L}{\Gamma^2} c_{\rm qel} 
\ln\left|\frac{\Gamma}{eV-\hbar \omega_0}\right|,
\end{equation}
with the dimensionless coefficient 
 \begin{equation} \label{cqel}
c_{\rm qel} = \sum_\alpha  \frac{-\alpha\frac{\Gamma_\alpha}{\Gamma}
\frac{\bar \mu-\alpha\omega_0/2}{\Gamma} (1+ [\bar\mu/\Gamma+\alpha \omega_0
/2\Gamma]^2)} { [1+(\bar \mu/\Gamma+\omega_0/2\Gamma)^2]^2
 [1+(\bar \mu/\Gamma-\omega_0/2\Gamma)^2]^2 }.
\end{equation}
It is worth mentioning that $c_{\rm qel}=0$ for large $\Gamma\gg \omega_0$
at the symmetric point $\Gamma_L=\Gamma_R$. 
Away from this limit, however,
the logarithmic term in Eq.~(\ref{qelcorr}) will be present.
All other contributions to $\delta I_{\rm qel}$ beyond Eq.~(\ref{qelcorr}) 
are smooth and featureless 
at $eV\approx \hbar \omega_0$, and do not affect the characteristic
feature in $d^2 I / dV^2$, whereas the singular contribution (\ref{qelcorr}) 
is logarithmically divergent at the threshold.
Note that this logarithmic divergence due to quasi-elastic processes 
creates a symmetric dip or peak (depending on the sign of $c_{\rm qel}$) 
in the differential conductance at $eV=\hbar
\omega_0$, while the inelastic contributions are responsible for
a step feature.  In the full $d^2 I / dV^2$ curve, this translates to
{\sl asymmetric} dips or peaks.
The relative importance of inelastic versus quasi-elastic contributions
can be judged from the ratio
 $c_{\rm inel}/c_{\rm qel}$.
For the symmetric case, $\Gamma_L=\Gamma_R=\Gamma/2$, a simple
result follows from Eqs.~(\ref{cdef}) and (\ref{cqel}),  
\begin{equation}
 r=  \left| \frac{c_{\rm inel}}{c_{\rm qel}} \right|= 
\frac{|\Gamma^2+ \bar\mu^2 - \omega_0^2/4|}
{\Gamma \omega_0/2}.
\end{equation}
For small $\omega_0/\Gamma$ or large $\bar \mu$, 
we have $r\gg 1$ and the inelastic channel always
dominates, while for large $\omega_0$, quasielastic
processes can be more important.
The perturbative results (\ref{qelcorr}) and (\ref{step}) obviously break down
close to the threshold voltage. At $T=0$ and in the 
absence of an external bath, the only way to account for 
the finite lifetime of the phonon, and hence the smearing
of the step and/or peak features, is to take into account the 
electronic polarization in the phonon GF. 
The retarded polarization function $\chi^r(\omega)$ will then result in a 
damping $\gamma\simeq - g^2\chi^r_I(\omega_0)$ of the phonon mode.
We obtain after some algebra the nonequilibrium 
electronic polarization function in analytical form.
In the particle-hole symmetric case, this result simplifies to  
\begin{equation}
\chi^r(\omega)=\frac{\Gamma}{\pi} \frac{1}{\omega(\omega+2i\Gamma)}
\ln\left(1-\frac{\omega(\omega+2i\Gamma)}{\Gamma^2+V^2/4}\right) .
\end{equation}
This implies the estimate $\gamma\simeq g^2\omega_0/\pi \Gamma^2$
in the limit of a soft phonon $\omega_0\ll\Gamma$, and 
$\gamma\simeq g^2\Gamma/\omega_0^2$ for a hard phonon $\omega_0\gg\Gamma$.
However, phonon damping is, in fact, a higher-order effect in
the electron-phonon coupling, and to consistently account for the
finite damping $\gamma$ while keeping current conservation intact
remains a theoretical challenge. Other effects of higher-order
diagrams include the proliferation of steps and/or peaks
at multiples of the phonon frequency $\omega_0$. Indeed, 
it is easy to see that the $g^{2n}$th-order rainbow diagram in the 
electronic self-energy produces a step feature in the
differential conductance at the voltage $V=n\hbar\omega_0/e$. 
The appearance of such step features at multiples of $\omega_0$
is closely related to the strong-coupling picture obtained
through a polaron transformation.\cite{brat}
However, when going beyond the lowest order in $g$,
vertex corrections are also expected to be important.
Unfortunately, the proper treatment of such 
nonequilibrium many-body effects 
(respecting the requirements posed by current conservation) 
remains a challenging task and is beyond the scope of this work. 
For weak electron-phonon coupling, which appears 
to be appropriate for many experiments, none of these fine details matter 
in any case, and the $g^2$ calculation is sufficient.
The $d^2 I/dV^2$ feature is then 
determined by Eqs.~(\ref{step}) and (\ref{qelcorr}),
where the damping $\gamma$ in the phonon mode acts to broaden
the delta function in Eq.~(\ref{step}) within a phenomenological description.

The above results allow us to clarify the question of peak vs dip in the 
second derivative, $d^2  I/dV^2$, which arises due to
the singular inelastic  correction (\ref{step}).
For $\Gamma_L=\Gamma_R$ and $\bar\mu=0$, where the transparency (\ref{trans})
is ideal, ${\cal T}=1$, we observe from Eq.~(\ref{cdef}) that 
for $\omega_0>2\Gamma$, instead of the expected dip ($c_{\rm inel}>0$),
one actually observes a peak.  For $\bar\mu\ne 0$, once 
$|\bar\mu|>\sqrt{\Gamma^2+\omega_0^2/4}$, one finds
a peak.  Note that for $\bar\mu=\pm \Gamma$, the transparency
(\ref{trans}) is precisely $1/2$, thereby allowing us to 
rationalize why previous numerical studies 
for related models\cite{paulsson,alfredo} reported  a 
${\cal T}=1/2$ criterion for the transition from peak
to dip.  This value correctly describes the transition in
the limit of a soft phonon, $\omega_0\ll \Gamma$,  and assuming
symmetric contacts, $\Gamma_L= \Gamma_R$.  The value ${\cal T}=1/2$
was, in fact, established precisely in this parameter 
regime.\cite{paulsson,alfredo}
 Our analytical result (\ref{cdef}),
shows, however, that, in general, the {\sl transition 
is nonuniversal and determined by all parameters}.
For example, it can be achieved either
by tuning ${\cal T}$ --- where the precise transition value depends
also on $\omega_0$ and the asymmetry $\Gamma_L-\Gamma_R$, 
and is only approximately given by ${\cal T}=1/2$ ---
or by changing other parameters, such as $\Gamma_R-\Gamma_L$
or the ratio between phonon frequency and hybridization, $\omega_0/\Gamma$.  
The nonuniversality of the step is also implicit in the 
early work on phonon-assisted tunneling through a 
resonant level by Glazman and Shekhter.\cite{gs}
On top of this peak or dip structure due to inelastic processes,
the quasielastic contribution causes a singular
response near the threshold value $eV=\hbar \omega_0$.
This logarithmic correction to the differential conductance 
implies an asymmetric line shape in $d^2 I/dV^2$ as discussed above.  
Such asymmetries have frequently been reported experimentally and
in numerical calculations,\cite{nitzan2} and they are a direct consequence
of the Engelsberg-Schrieffer singularity. 

We thank A. Zazunov and T. Novotny for useful discussions.
This work was supported by the DFG SFB TR 12 and by the
ESF network INSTANS. A.O.G. thanks the Humboldt Foundation
for a Friedrich-Wilhelm Bessel grant enabling his extended stay
in D\"usseldorf.

\end{document}